\documentclass[%
 reprint,
superscriptaddress,
nofootinbib,
 amsmath,amssymb,
 aps,showpacs,
pra
]{revtex4-1}

\usepackage{graphicx}
\usepackage{dcolumn}
\usepackage{bm}

\usepackage{dsfont}

\usepackage[T1]{fontenc}

\usepackage[percent]{overpic}	
\usepackage{color}		
\usepackage{rotating}
\usepackage{placeins} 

\begin{document}

\title[Many-body processes in black and grey matter-wave solitons]
{Many-body processes in black and grey matter-wave solitons}
\author{Sven Kr\"onke}
  \email{skroenke@physnet.uni-hamburg.de}
  \affiliation{Zentrum f\"{u}r Optische Quantentechnologien,
  Universit\"{a}t Hamburg, Luruper Chaussee 149, 22761 Hamburg, Germany}
\author{Peter Schmelcher}
  \email{pschmelc@physnet.uni-hamburg.de}
  \affiliation{Zentrum f\"{u}r Optische Quantentechnologien,
  Universit\"{a}t Hamburg, Luruper Chaussee 149, 22761 Hamburg, Germany}
  \affiliation{The Hamburg Centre for Ultrafast Imaging,
  Luruper Chaussee 149, 22761 Hamburg, Germany}

\begin{abstract}
We perform a comparative beyond mean-field study 
of black and grey solitonic excitations in a finite ensemble of ultracold bosons
confined
to a one-dimensional box. An optimized density-engineering potential is
developed and employed together with phase-imprinting to cleanly initialize
grey
solitons. Based on our recently developed Multi-Layer Multi-Configuration
Time-Dependent Hartree Method for Bosons, we demonstrate that quantum fluctuations 
limit the lifetime of the soliton contrast, which prolongs with 
increasing soliton velocity. A natural orbital analysis reveals a two-stage 
process underlying the decay of the soliton contrast. The broken parity symmetry
of grey solitons results in a local asymmetry of the orbital mainly responsible
for the decay, which leads to a characteristic asymmetry of remarkably localized
two-body correlations. The emergence and decay of these correlations
as well as their displacement from the instantaneous soliton position 
are analysed in detail. Finally, the role of phase-imprinting for the
many-body dynamics is illuminated and additional non-local correlations
in pairs of counter-propagating grey solitons are unravelled.
\end{abstract}

\pacs{03.75.Lm, 67.85.De}

\maketitle

\section{Introduction}\label{sec_intro}
Solitons are very peculiar solutions of non-linear wave equations
emerging in various fields of physics ranging from non-linear optics
to shallow water waves \cite{waves_called_solitons,physics_of_solitons}. 
These solutions are characterized by their form-stability under time evolution
and even under collisions so that solitons can behave akin to classical
particles and may be
described as such under certain conditions. In
particular, ultracold bosonic quantum gases allow both theoretically
and experimentally for thorough investigations of
dark solitons, i.e. localized density minima accompanied by characteristic
phase jumps in the order parameter, residing on the background density profile of
trapped atomic clouds
(\cite{emerg_phenom_BEC_08,dark_solitons_in_atomic_BEC_Frantzeskakis_JPA2010}
and
ref. therein). The fully integrable Gross-Pitaevskii mean-field equation for a
quasi-one-dimensional 
uniform, perfect Bose-Einstein condensate, for instance, posseses a dark soliton
solution \cite{tsuzuki_nonlinear_1971,
Bose-Einstein_Condensation_Dilute_Gases_Pethick_Smith_2008,emerg_phenom_BEC_08,
dark_solitons_in_atomic_BEC_Frantzeskakis_JPA2010}
completely characterized by the ratio $\beta$ of soliton velocity $u$ to the
speed of
sound $s=\sqrt{\rho_0 g/m}$:
\begin{eqnarray}\label{grey_sol_analy}
\psi_\beta(x,t)=&\phantom{x}&\sqrt{\rho_0}e^{-i\mu_0 t/\hbar}\;\times\\\nonumber
&\phantom{x}&\times\;\left(i\beta +\eta\tanh\Big(\eta\frac{x-\beta st
-x_0}{\xi} \Big)\right),
\end{eqnarray}
with $|\beta|\leq 1$, $\eta=\sqrt{1-\beta^2}$, bulk density $\rho_0$, 
chemical potential $\mu_0=\rho_0g$, healing
length $\xi=\hbar / \sqrt{m\rho_0g}$, atomic mass $m$ and contact interaction
strength $g$. Black solitons ($\beta=0$) do not move and feature a density notch
with a phase jump of $\pi$, while grey
solitons ($|\beta|>0$) are moving objects with a finite density minimum and a
smaller phase jump. 

As excited states, however, dark solitons may suffer from 
various sources of instabilities ranging from thermodynamic
(e.g.
\cite{dynamics_ds_elongated_BECs_Shlyapnikov_Senkstock_Lewenstein_PRL2002,
gangardt_quantum_2010,cockburn_fluctuating_2011} and ref. therein) and
dynamical instabilities  
(e.g.\
\cite{dynamics_ds_elongated_BECs_Shlyapnikov_Senkstock_Lewenstein_PRL2002,
brand_solitonic_2002,becker_inelastic_2013,ku_motion_2014,
donadello_observation_2014,
mateo_chladni_2014} and ref. therein) to decay as a consequence of
quantum fluctuations. Since nowadays
experiments can be operated at effectively zero temperature and with a
high aspect ratio of the transverse and longitudinal traps,
ultracold bosonic quantum gases serve as ideal systems for exploring
the quantum nature and correlation effects in solitonic excitations. 
The number of atoms and the interaction strength obviously constitute key
parameters, which determine the intensity of correlations and are, most
importantly,
controllable in nowadays experiments.
In passing, we note that also the illumination of beyond mean-field
effects in vortex excitations has recently attracted interest
\cite{vortex_nucleation,weiner_angular_2014,tsatsos_vortex_2014,
wells_vortex_2014}. 

Usually, the form-stability of the dark solitons is regarded as a compensation
of dispersion by the non-linearity of the Gross-Pitaevskii equation
\cite{tsuzuki_nonlinear_1971}. The actual
many-body Schr\"odinger equation, however, is linear and should also be able
to describe solitons in appropriate parameter regimes. The ongoing theoretical
efforts from this linear perspective can be classified into two directions:
The deductive approach 
\cite{kulish_comparison_1976,ishikawa_solitons_1980,kanamoto_metastable_2010,
sato_exact_2012,sato_quantum_2012,wadkin-snaith_quantum_2012,
astrakharchik_liebs_2013,draxler_particles_2013,karpiuk_correspondence_2014}
aims at
establishing a relation between the hole-like
type II excitations 
of the solvable Lieb-Liniger model
\cite{exact_analysis_interacting_Bose_gas_I_Lieb_Liniger,exact_analysis_interacting_Bose_gas_II_Lieb_Liniger}
and the Gross-Pitaevskii soliton solutions
\cite{tsuzuki_nonlinear_1971}. In contrast to this,
we follow the inductive approach in which one either starts with a
mean-field product state featuring a soliton or uses experimentally relevant
protocols to prepare a many-body state resembling
the properties of a dark soliton.
The subsequent time-evolution is then studied with beyond mean-field
methods.

Shortly after the first experimental implementation of dark solitons in
Bose-Einstein condensates \cite{dark_solitons_in_bec_Burger_Sengstock_PRL1999},
the inductive
approach studies predicted a dynamical instability due to quantum fluctuations
\cite{quantum_depletion_of_excited_condensate_Sacha_Dziarmaga_Karkuszewski_PRA2002,
quantum_depl_dark_sol_anomalous_mode2002,
law_quantum_2002,images_of_dark_soliton_in_depleted_condensate_Dziarmaga_Karkuszewski_Sacha,
law_dynamic_2003,images_BEC_diag_dyn_Bogoliubov_vacuum_Dziarmaga_Sacha,
dziarmaga_quantum_2004,
quantum_entangled_dark_solitons_in_optical_lattices_Carr_PRL09,
quantum_mb_dynamics_of_dark_solitons_in_optical_lattices_Carr_PRA09,
martin_quantum_2010,delande_many-body_2014}: 
The density minimum of a dark soliton is incoherently filled with atoms on
potentially experimentally relevant time scales. On the one hand side, this
dynamical quantum depletion effect has been studied within the Bogoliubov
perturbation theory
\cite{quantum_depl_dark_sol_anomalous_mode2002,
law_quantum_2002,images_of_dark_soliton_in_depleted_condensate_Dziarmaga_Karkuszewski_Sacha,
law_dynamic_2003,images_BEC_diag_dyn_Bogoliubov_vacuum_Dziarmaga_Sacha} as well
as a non-perturbative variant \cite{dziarmaga_quantum_2004}
highlighting the role of the localized zero (anomalous) mode in uniform (trapped) systems
as main contributor to the filling of the density minimum. On the other hand,
the numerically exact time-evolving block-decimation technique (TEBD) has
been employed in finite lattices
\cite{quantum_entangled_dark_solitons_in_optical_lattices_Carr_PRL09,
quantum_mb_dynamics_of_dark_solitons_in_optical_lattices_Carr_PRA09}
and continuous systems within the Bose-Hubbard approximation
\cite{delande_many-body_2014}. As the main experimental signatures, the
relaxation of the reduced one-body density to a flat profile and a quantum
fluctuation induced inelasticity of a binary soliton collision
were reported.
Moreover, at times when the reduced one-body density, i.e. the average over
many single shot measurements, has already relaxed to a flat distribution,
the histograms of simulated destructive $N$-atom single shot measurements
have revealed a soliton-like density
minimum at random positions \cite{
images_of_dark_soliton_in_depleted_condensate_Dziarmaga_Karkuszewski_Sacha,
images_BEC_diag_dyn_Bogoliubov_vacuum_Dziarmaga_Sacha,
comment_on_Carr_PRL_PRA_Sacha_PRL2010,
reply_on_comment_on_Carr_PRL_PRA_Carr_PRL2010,
delande_many-body_2014}. 
These findings suggest the existence of
highly non-trivial correlations being unravelled in single shot 
measurements.

The above works have almost exclusively focussed on black solitons and,
except for some side aspects of
\cite{wadkin-snaith_quantum_2012,dziarmaga_quantum_2004,
quantum_entangled_dark_solitons_in_optical_lattices_Carr_PRL09,
quantum_mb_dynamics_of_dark_solitons_in_optical_lattices_Carr_PRA09}, grey
solitons have not been studied beyond the mean-field approximation so far.
Therefore, this work aims at a systematic comparison of beyond mean-field
signatures in black solitons and their experimentally more relevant grey counterparts. 
Having introduced our setup and the employed {\it ab initio} method in
Sect.\ \ref{ssec_setup} and \ref{ssec_method}, respectively, we present a
semi-analytically optimized density-engineering scheme, which allows, in
combination with a phase-imprinting procedure, to robustly and cleanly generate
grey solitons (Sect.\ \ref{ssec_initial_state}). Although the broken parity
symmetry of a grey soliton results in a larger variety of allowed 
incoherent scattering channels compared to a black soliton of well defined
parity, grey
solitons prove to be more stable in terms of a longer contrast lifetime (Sect.
\ref{ssec_density})
and slower dynamical quantum depletion (Sect.\ \ref{ssec_dep_norb}). Despite of
the extensive literature on many-body effects in black solitons, the evolution
of local two-body correlations, which are experimentally accessible via
density-density correlation measurements, has not 
been investigated so far. In Sect.\ \ref{sec_g2}, we explore
the occurrence of spatially well localized bunching and antibunching
correlations for dark solitons: 
Whereas
these correlations are symmetrically arranged around a black soliton of well
defined parity,
the localized bunching correlations become displaced to the back of a grey
soliton
with subsonic velocity while emerging. The characteristic asymmetric
correlation pattern of a grey soliton is traced back to a local asymmetry of
the single-particle state most responsible for the soliton decay. In Sect.
\ref{sec_no_impr}, the role of the phase-imprinting is illuminated,
demonstrating that imposing a phase profile accelerates the quantum decay
while omitting this procedure results in pairs of counter-propagating,
long-living grey solitons. Besides the localized two-body correlations
identified for a single grey soliton, we observe additional non-local
correlations in the
latter case. We conclude in Sect.\ \ref{sec_concl}.
\section{Setup, computational method and initial state
preparation}\label{sec_setup_method}

\subsection{Setup}\label{ssec_setup}
In this work, we study the dynamics of $N$ indistinguishable bosons in a
one-dimensional box potential of length $L$. Such potentials are realizable via
crossed optical dipole or strong transverse lattice potentials
combined with various implementation techniques for box potentials in the 
longitudinal direction
\cite{meyrath_bose-einstein_2005,henderson_experimental_2009,
gaunt_bose-einstein_2013}. Considering bosonic atoms interacting 
solely via the contact interaction such as $^{87}$Rb at so low temperatures that
both $k_BT$ and the chemical potential $\mu$ are much smaller than the
excitation energy $\hbar \omega_\perp$ of the axially symmetric
transverse harmonic potential justifies to consider
a purely one-dimensional model,
\begin{equation}
\hat H = \sum_{i=1}^N \frac{\hat p_i^2}{2m} + g\sum_{1\leq i<j\leq N}
	\delta(\hat x_i-\hat x_j)
\end{equation}
with hard wall boundary conditions. Here, $m$ denotes the mass of an atom and
$g$ relates to the three-dimensional s-wave scattering length $a_s$
and the transverse trapping potential via
$g=\frac{4\hbar^2}{m}\frac{a_s}{a_\perp^2} / (1-C\frac{a_s}{a_\perp})$
with $a_\perp=\sqrt{2\hbar/(m\omega_\perp)}$ and a numerical constant $C$ 
\cite{Olshanii_PRL98_quasi_1d_scattering}.
In the thermodynamic limit, the interaction regime of our system is fully
characterized by the coupling constant
$g$ and the linear atom density $\rho_0=N/L$ in terms of the
dimensionless Lieb-Liniger parameter $\gamma=mg/(\hbar^2\rho_0)$ measuring the
ratio of interaction and kinetic energy \cite{exact_analysis_interacting_Bose_gas_I_Lieb_Liniger,exact_analysis_interacting_Bose_gas_II_Lieb_Liniger}. 
Violating the above prerequisites for a quasi-one-dimensional description
gives rise to complex dynamical instabilities of dark solitons, which
is of current experimental and theoretical interest
\cite{becker_inelastic_2013,ku_motion_2014,donadello_observation_2014,
mateo_chladni_2014} but goes beyond the
scope of this work.

Our system features three length scales: The mean inter-particle distance 
$\rho_0^{-1}$, the condensate healing length 
$\xi$ and the box length $L$. In order to resemble the properties
of a mean-field soliton in the initial state, we focus on weak interactions.
Moreover, we aim at separating the length scale of the soliton, $\xi$, well
from the box length. Thus,
we initially operate in the Thomas-Fermi mean-field regime, whose
validity range for the ground state (in the thermodynamic limit) is given by
$N^{-2}\ll \gamma\ll 1$ or $\rho_0^{-1}\ll \xi\ll L$ \cite{lieb_one-dimensional_2003}. Here, 
the first (second) inequality ensures the applicability of the mean-field 
(Thomas-Fermi) approximation. In the following, $N=100$ bosons
with $\gamma=0.04$ and $L=20\xi$ are considered.
We remark that the above considerations apply only to ground states in
the thermodynamic limit and therefore 
do not exclude dynamical quantum depletion in the
many-body quantum dynamics of our finite ensemble.

In the following, we use the healing length based unit system, i.e. 
$\xi$ as the length and $\mu_0$ as the energy unit implying
that the correlation time $\tau=\xi/s=\hbar/\mu_0$ serves as the time unit.
The dimensionless Hamiltonian reads
$\hat H' = \sum_i {{\hat p}_i}^{\prime2}/2+g'\sum_{i<j}\delta(\hat x'_i-\hat x'_j)$
with $g'=\sqrt{\gamma}$. For simplicity, we will omit the dash in
the notation from now on.

\subsection{Computational method}\label{ssec_method}
Going non-perturbatively beyond 
the mean-field approximation requires the usage of sophisticated many-body
methods such as e.g.\ TEBD 
\cite{eff_class_simulation_of_slightly_entangled_quantum_comp_VidalPRL2003} 
in order to soften the exponential increase of complexity with the number of degrees
of freedom. In this work, we employ the 
recently developed Multi-Layer Multi-Configuration Time-Dependent 
Hartree Method for Bosons (ML-MCTDHB) \cite{kronke_non-equilibrium_2013,cao_multi-layer_2013},
which is a flexible, 
{\it ab initio} method for solving the time-dependent 
Schr\"odinger equation for bosons or bosonic mixtures in one- or higher dimensions.
This method rests on expanding the total many-body wave function with respect
to a variationally optimized time-dependent basis, which
spans the optimal subspace of the Hilbert space at each
instant in time and thereby reduces the necessary basis size significantly compared
to an expansion w.r.t. a time-independent basis.

When applied to 
a single bosonic species only, ML-MCTDHB reduces to the 
Multi-Configuration Time-Dependent Hartree Method for Bosons (MCTDHB), which was
firstly introduced in 
\cite{role_of_excited_states_in_splitting_of_BEC_by_time_dep_barrier_Steltsov_PRL2007,
MCTDHB_many_body_dynacmics_of_bosonic_systems_phys_rev_a_Alon_Streltsov_Cederbaum}.
In this case, the total wave function $|\Psi_t\rangle$ is expanded with respect to bosonic 
number states $|n_1,...,n_M\rangle_t$ being based on time-dependent
single-particle functions (SPFs) $|\phi_i(t)\rangle$, $i=1,...,M$:
$|\Psi_t\rangle=\sum_{\vec n|N} A_{n_1,...,n_M}(t) |n_1,...,n_M\rangle_t$,
where $\vec n=(n_1,...,n_M)$ encodes the occupation numbers of the $M$ SPFs
and $\vec n|N$ indicates that the summation runs over {\it all} occupation
numbers $n_i$ that sum up to $N$. The (ML-)MCTDHB equations of motion
for the expansion coefficients $A_{n_1,...,n_M}(t)$ and the SPFs $|\phi_i(t)\rangle$
are then derived from a variational principle.
In this way, the method provides a variationally optimized 
$|\Psi_t\rangle$ within this class of trial wave functions characterized by the
given 
number of SPFs $M$. It can be proven that the coupled (ML-)MCTDHB 
equations of motion respect certain symmetries
such as parity \cite{cao_multi-layer_2013}, which will become of importance for this work. 
Varying $M$ allows to go from the mean-field
limit ($M=1$) to the w.r.t. to the given spatial grid numerically exact limit,
in which $M$ equals the number of
grid points $n$.
In \ref{app_convergence}, a convergence discussion for our simulation data is
provided.
In view
of the ansatz for $|\Psi_t\rangle$, it becomes inevitable to specify a 
recipe for initializing the {\it many-body} wave function $|\Psi_{t=0}\rangle$
featuring a dark soliton.
\subsection{Initial state preparation}\label{ssec_initial_state}
For a given number of SPFs $M$, the objective of our initialization recipe is to 
prepare a many-body state which features only little depletion and the density
and phase characteristics of a dark soliton in the dominant natural orbital, 
i.e. the eigenvector of the reduced one-body density operator $\hat \rho_1$
with the largest eigenvalue (natural population). In particular, we aim at 
generating a clean solitonic excitation, which is favourable for both 
clear physical insights and the convergence of our numerical method. For these
reasons, the phase- and density-engineering scheme is applied
\cite{dark_soliton_creation_in_BECs_CarrPRA2001}: Via imaginary time
propagation of the ML-MCTDHB equations of motion, an initial guess
for the many-body wave function is relaxed to the ground state of the box
potential with an additional localized barrier $V(x)$ of such shape
that the induced density minimum resembles the density profile of a dark
soliton.
After switching off $V(x)$ instantaneously, an intense laser-induced potential 
$\theta(x)/T$ is applied for a duration $T\ll \tau$, where $\theta(x)$ shall 
coincide
with the phase-profile of a dark soliton (\ref{grey_sol_analy}). Assuming that 
this potential dominates all other terms in the Hamiltonian, the corresponding
time-evolution operator reads 
$\hat U_T\approx \bigotimes_{j=1}^N e^{-i\theta(\hat x_j)}$. Its action on 
the total wave function is obviously equivalent to an instantaneous replacement 
$|\phi_i(0)\rangle\rightarrow e^{-i\theta(\hat x)}|\phi_i(0)\rangle$,
$i=1,...,M$.

For generating a black soliton at the box centre $x_0=0$, we follow the strategy
of \cite{quantum_entangled_dark_solitons_in_optical_lattices_Carr_PRL09,
quantum_mb_dynamics_of_dark_solitons_in_optical_lattices_Carr_PRA09,
delande_many-body_2014} by using a Gaussian barrier $V(x)=h\exp(-x^2/(2w^2)$
with $h=60\mu_0$ and $w\approx0.07\xi$. In principle, it is also possible to
generate a grey soliton 
by means of a Gaussian barrier
fine tuned to an appropriate e.g.\ $\tanh$-phase profile
\cite{dark_soliton_creation_in_BECs_CarrPRA2001}. We,
however, use the analytical phase profile of (\ref{grey_sol_analy}) for
$\theta(x)$ and develop a semi-analytical formula for an optimized
density-engineering potential: Having determined the Gross-Pitaevskii
ground state $\Phi(x)$ of the box numerically, we define a real-valued target
Gross-Pitaevskii orbital 
$\psi^\beta_{\rm target}(x)\propto |\Phi(x)\psi_\beta(x,0)|$, which we assume to
be (i) twice differentiable in the box domain $(-L/2,L/2)$ and
(ii) to solve a stationary Gross-Pitaevskii equation with an unknown potential
$V_\beta(x)$. Then we may express $V_\beta(x)$ in terms of $\psi^\beta_{\rm
target}(x)$,
\begin{eqnarray}\label{dens_eng_pot}
V_\beta(x)={\rm const.}&+&\frac{1}{2}
\frac{\partial_x^2\psi^\beta_{\rm target}(x)}{
\psi^\beta_{\rm target}(x)+f(\epsilon,\psi^\beta_{\rm target}(x))}\\\nonumber
	&-&g(N-1)|\psi^\beta_{\rm target}(x)|^2,
\end{eqnarray}
where $f(\epsilon,z)=\epsilon\exp(-|z|/\epsilon)$, $\epsilon\ll1$
regularizes\footnote{In our applications, we can safely 
omit the regularization
as long as $|\beta|>0$. For $\beta=0$, however, (\ref{dens_eng_pot}) becomes
ill-defined since $\psi^0_{\rm target}(x)$ is not differentiable at
$x=x_0$. In fact,
the (not regularized) potential maximum $V_\beta(x_0)$ diverges as
$1/\beta^2$ for $\beta\rightarrow 0$ while the full width half maximum of
this potential remains finite.}
possible zeros or too small values of $\psi^\beta_{\rm target}(x)$. By
construction, the Gross-Pitaevskii ground state of the box plus $V_\beta(x)$
potential equals $\psi^\beta_{\rm target}(x)$ and due to weak interactions we
may expect that a corresponding many-body calculation beyond the mean-field
approximation results in a one-body density $\rho_1(x)\approx|\psi^\beta_{\rm
target}(x)|^2$.
This scheme turns out to be both very robust and practicable as it does not
require any optimization algorithm. As long as the target state is twice
differentiable (which is related to the absence of zeros) and features a length
scale separation required by the local density approximation (LDA), this
approach allows to cleanly generate grey solitons while only marginally exciting other
modes. For demonstrating its versatility, we have also successfully applied
this recipe to initialize oscillating dark
solitons in harmonic traps (plots not shown). Moreover, if one can increase the
healing length
$\xi$ above the optical diffraction limit by means of e.g.\ a Feshbach
resonance,
this density-engineering scheme might also improve the initial state
preparation in experiments. 
\section{Results}\label{sec_results}
We begin our comparative study of black and grey solitons beyond the
mean-field approximation with the reduced one-body density
(Sect.\ \ref{ssec_density}) and the depletion (Sect.\ \ref{ssec_dep_norb}).
Then, the reduced one-body dynamics is unravelled in terms of a natural
orbital analysis, characterizing the single-particle state mostly responsible
for the soliton decay. Concerning black solitons, our results are fully
consistent with TEBD simulations of discrete
\cite{
quantum_entangled_dark_solitons_in_optical_lattices_Carr_PRL09,
quantum_mb_dynamics_of_dark_solitons_in_optical_lattices_Carr_PRA09} and
continuous systems \cite{delande_many-body_2014}.
Afterwards, we explore the evolution of local
two-body correlations (Sect.\ \ref{sec_g2}). To the best of our knowledge,
the dynamics of two-body correlations has not been studied so far, not even for
black solitons. Finally, the role of the
phase-imprinting procedure is illuminated (Sect.\ \ref{sec_no_impr}).
\begin{figure}[t]
 \centering
\includegraphics[width=0.4\textwidth]{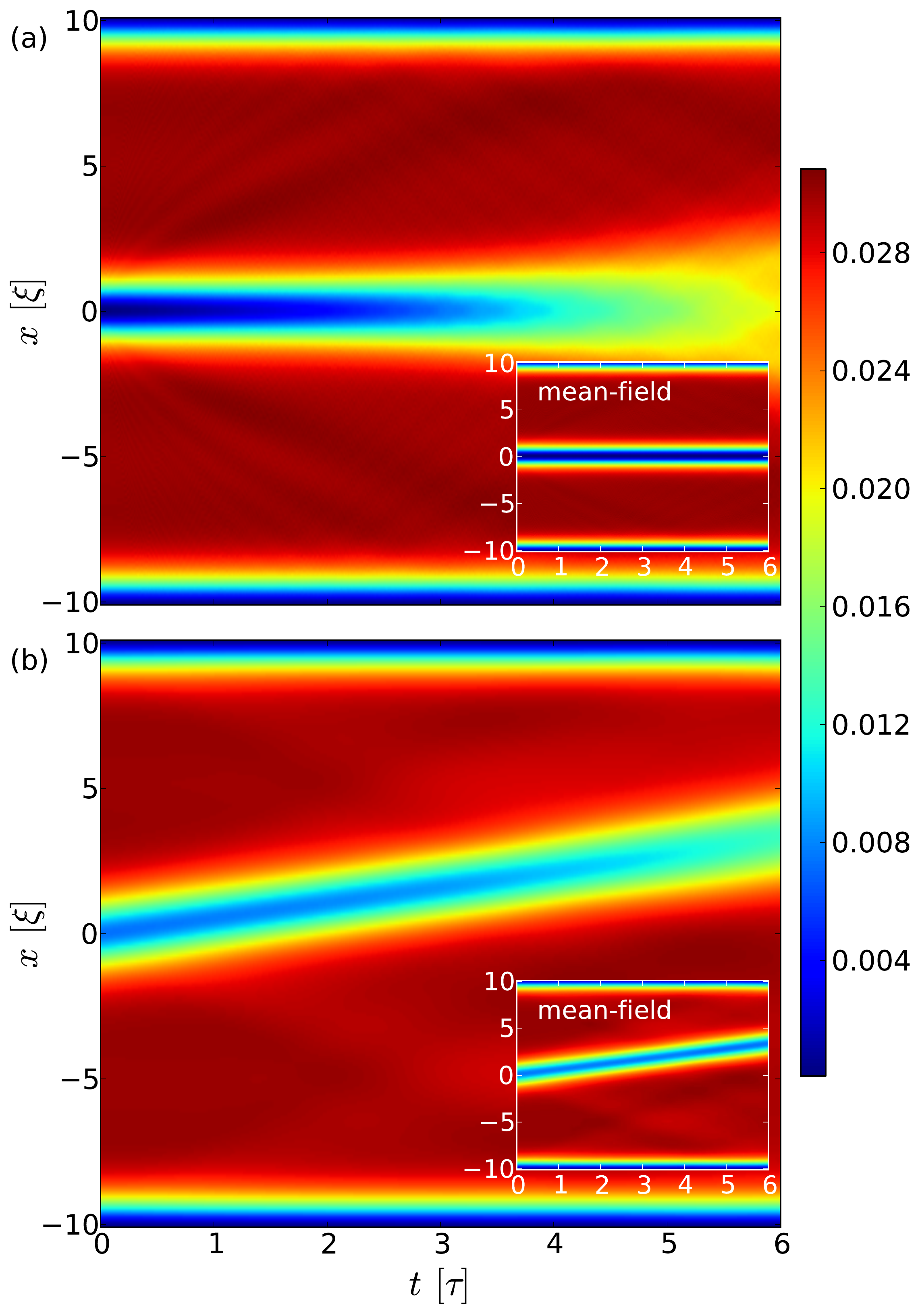}
\caption{Time-evolution of the reduced one-body density $\rho_1(x;t)$ for 
$N=100$ bosons in a box of length $L=20\xi$ and a Lieb-Liniger parameter
$\gamma=0.04$.
The initial state resembles the properties of a black ($\beta=0.0$), a grey
soliton 
($\beta=0.5$)
in (a) and (b), respectively. The calculations are performed with
$M=4$ optimized SPFs. Insets show $M=1$ mean-field simulations.
.}
\label{fig_density}
\end{figure}
\subsection{Reduced one-body density and contrast}\label{ssec_density}
All single-particle properties are described by the reduced one-body
density operator $\hat \rho_1(t)={\rm tr}_1|\Psi_t\rangle\!\langle \Psi_t|$
being obtained by tracing out all bosons but a single one in the density
operator of the $N$-body system. Firstly, let us compare the evolution of the 
reduced one-body density $\rho_1(x;t)=\langle x|\hat \rho_1(t)|x\rangle$
for a black and grey soliton. In Fig.\ \ref{fig_density}, we present both
mean-field
and ML-MCTDHB calculations performed with $M=4$ optimized SPFs, to which we will
refer as many-body simulations in the following as in contrast to the 
effective single-particle 
mean-field theory. One can clearly see 
that the applied density- and phase-engineering scheme generates stable solitons
within the mean-field approximation, whereas the density minimum is becoming 
filled with atoms in the many-body simulations. This filling process appears
to be slower in the case of a moving grey soliton compared to the black one.
In both the many-body and the mean-field simulations, one clearly notices 
that our optimized density-engineering potential (\ref{dens_eng_pot}) allows
for the generation of clean solitonic excitations: As a consequence of
employing a Gaussian barrier of finite width, the black soliton
 is accompanied by quite some short 
\begin{figure}[!htb]
 \centering
\includegraphics[width=0.42\textwidth]{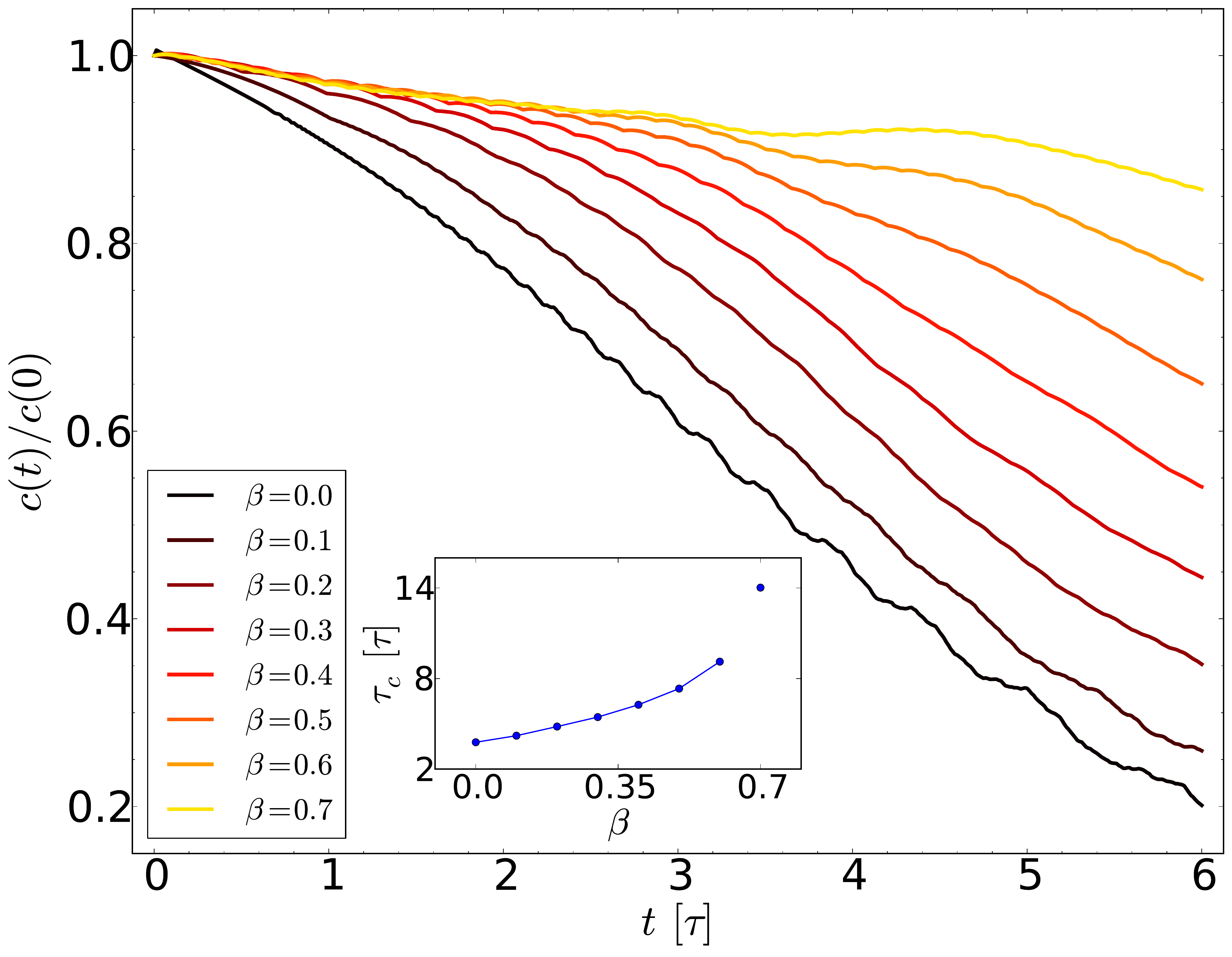}
\caption{Time-evolution of the relative contrast $c(t)/c(0)$
for 
dark solitons of various soliton velocity to speed of sound ratios $\beta=u/s$. Inset:
Contrast lifetime $\tau_c$ versus $\beta$.
All other parameters as in Fig.\ \ref{fig_density}.}
\label{fig_contrast}
\end{figure}
 wavelength phonons, which
 are visible as rays of density modulations in Fig.\ \ref{fig_density} (a). 
 In contrast to this,
only long wavelength phonon modes are marginally populated in the case of
the optimized grey soliton engineering (Fig.\ \ref{fig_density} (b)).

Next we quantify the lifetime of the soliton contrast,
\begin{equation}\label{constrast_def}
 c(t)=\frac{{\rm max}\,\rho_1(x,0)-\rho_1(x^s_t,t)}{{\rm
max}\,\rho_1(x,0)+\rho_1(x^s_t,t)},
\end{equation}
where $x^s_t$ refers to the soliton position at time $t$ being defined as 
the position of the density minimum. To compare results for black and grey
solitons,
we define the contrast lifetime $\tau_c$ to be the time after which
the relative contrast $c(t)/c(0)$ has dropped to $1/2$ (cf.\ also
\cite{quantum_entangled_dark_solitons_in_optical_lattices_Carr_PRL09}). As the
relative contrast is affected by some fluctuations due to phonons, we fit
$f(t)=a + bt^c$ with $c>0$ to
$c(t)/c(0)$ and
extract $\tau_c=[(\frac{1}{2}-a)/b]^\frac{1}{c}$.

From Fig.\ \ref{fig_contrast},
we may infer that the soliton contrast indeed lives the longer the closer
$\beta$ approaches unity, which is consistent with the analytical prediction 
in \cite{dziarmaga_quantum_2004}. Moreover, the decay of the grey soliton
relative contrast is approximately independent of $\beta$ for some time, which
turns out to be longer for larger $\beta$. After that time, the relative
contrast decays with a faster rate. 

The inset of Fig.\ \ref{fig_contrast}
depicts the contrast lifetime $\tau_c$ in dependence on $\beta$
showing that the lifetime of a grey soliton with $\beta=0.5$ is enhanced by
a factor of $1.9$ compared the black soliton. The data for $\beta=0.6$,
$\beta=0.7$
indicate that the lifetime can be increased much further. Yet we note that
the lifetime values for $0.4<\beta<0.7$ refer
to extrapolations to extend the converged results for a short time, while the
individual data point for
$\beta=0.7$ has been extrapolated because the soliton reaches the box boundary
before $\tau_c$. 

We finally remark that the dynamics of the relative contrast serves as a measure
for the quality of the density-engineering: While the relative contrast decay
is superimposed with only weak 	fluctuations for $0.1\leq \beta \leq 0.5$,
stronger perturbations are visible for the black soliton as well as $\beta
>0.5$.
The perturbations for large $\beta$ values, which take place
 on a relatively long time-scale of about $2\tau$, raise from
the fact that the soliton width separates less from the box length scale, which
undermines the LDA underlying eq. (\ref{dens_eng_pot}).
In the case of the black
soliton, the short-time fluctuations are a
consequence of
the Gaussian potential barrier being not of optimal shape and, moreover, 
the choice for its width and height being a compromise between resembling the
correct density profile and having a small condensate depletion at the same
time.
\subsection{Depletion and natural orbital analysis}\label{ssec_dep_norb}
\begin{figure}[t]
 \centering
\includegraphics[width=0.4\textwidth]{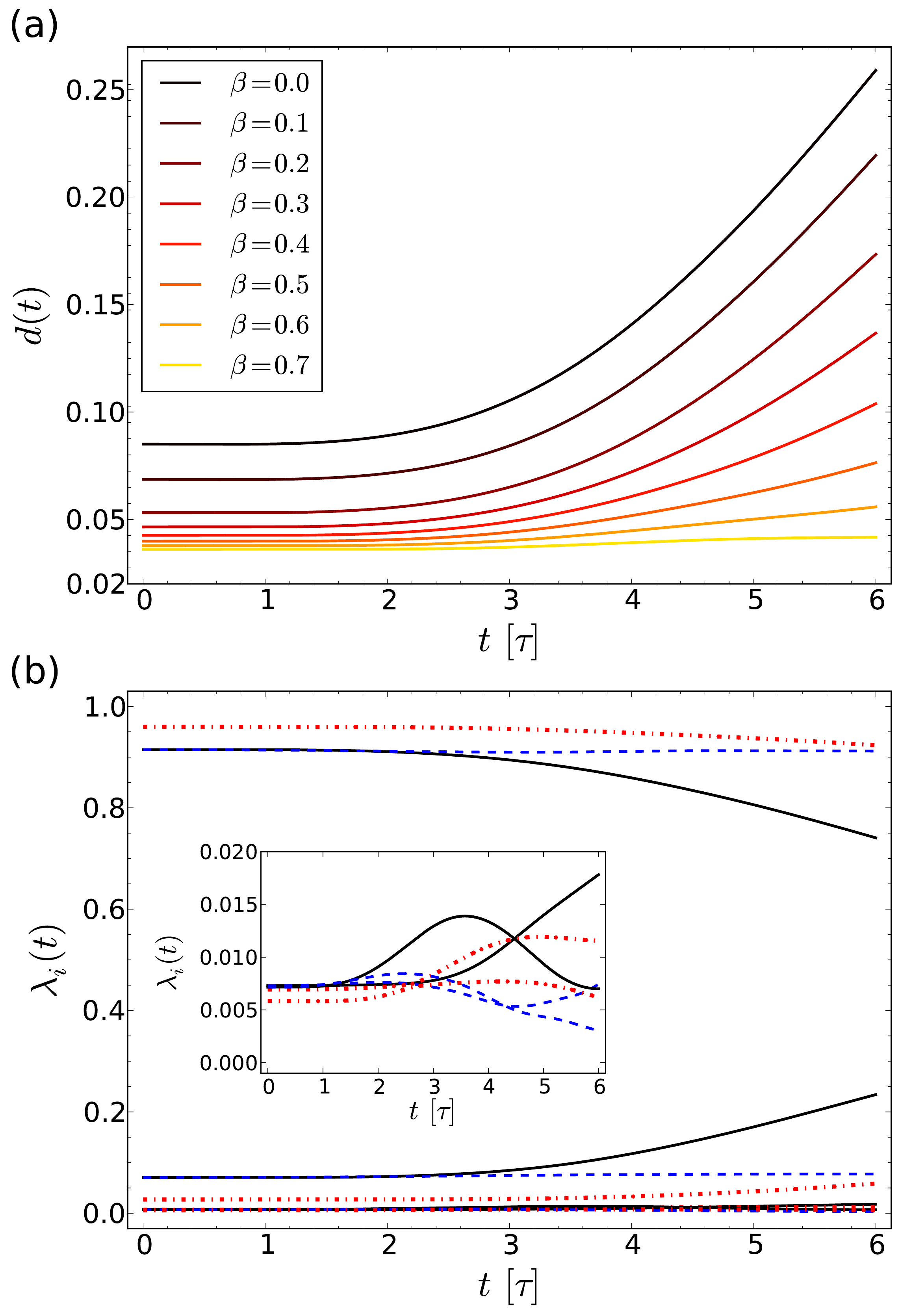}
\caption{(a) Time evolution of the depletion $d(t)$ for various 
$\beta$. (b) Evolution of the natural populations $\lambda_i$  for $\beta=0.0$ (solid black
lines) 
and $\beta=0.5$
(dashed dotted red lines). The blue dashed
lines refer to the natural populations for 
an $\beta=0.0$ density- but not phase-engineered
initial state. Inset: Close-up of the two least dominant natural populations. 
All other parameters as in Fig.\ \ref{fig_density}.}
\label{fig_depl_npop}
\end{figure}

In order to unravel the reduced one-body dynamics and to learn about the
structure of the many-body wave function, we inspect the spectral decomposition 
of the reduced one-body density operator,
\begin{equation}\label{rho1_spectral}
 \hat
\rho_1(t)=\sum_{i=1}^M\lambda_i(t)\,|\varphi_i(t)\rangle\!\langle\varphi_i(t)|,
\end{equation}
defining the natural orbitals
$|\varphi_i(t)\rangle$ and natural populations $\lambda_i(t)$
\cite{loewdin_norb55}.
First, we consider the depletion $d(t)=1-{\rm max}_i\,\lambda_i(t)\in[0,1]$
measuring how strongly the many-body wave function deviates from a perfect
Bose-Einstein condensed state \cite{Onsager_Penrose_BEC_liquid_He_PR_1956}.
In Fig.\ \ref{fig_depl_npop} (a), the evolution of $d(t)$ is compared for
various
values $\beta$. Initially, the maximal depletion of about $8.5\%$ is achieved
for 
$\beta=0.0$ and $d(0)$ decreases with increasing $\beta$. However, we may
also witness the impact of the $V_\beta(0)\sim1/\beta^2$ divergence as 
$\beta$ tends to zero
while the half-width-half-maximum of $V_\beta(x)$ saturates to a finite value:
As we decrease $\beta$ linearly but keeping it still finite, the depletion
$d(0)$ increases non-linearly\footnote{In fact, the data can be fitted well by
a sum of two exponentials with negative exponent coefficients.},
which is a consequence of $V_\beta(x)$
being optimal solely w.r.t. the (mean-field) density profile but
not w.r.t. beyond mean-field properties such as depletion.
Nevertheless, the initial many-body state is close enough to a perfectly
condensed state for all considered $\beta$ values in order to 
initially resemble
the important properties of mean-field dark solitons as we shall see below.

The subsequent dynamics of the depletion can be divided into two stages:
Firstly, $d(t)$ stays quite constant for some time, which turns out to
be the longer the larger $\beta$ is. Afterwards, a steep increase is followed
with a slope increasing with decreasing $\beta$. Thereby, we observe that
the greyer the soliton is the longer does it resemble mean-field
characteristics such as conservation of contrast and low depletion.

In Fig.\ \ref{fig_depl_npop} (b), we present the natural populations
$\lambda_i(t)$ for $\beta=0.0$ and $\beta=0.5$ as well as for a density-
but not phase-engineered initial state, which is discussed in Sect.
\ref{sec_no_impr}. In principle, $M=4$ natural orbitals are available in our
simulations, yet only two of them essentially contribute to the reduced
one-body density operator and thereby to the total many-body wave function. The
remaining two natural orbitals acquire natural populations of below $2\%$.
Again, the dynamics consists of two stages: Firstly,
the natural populations are stationary for a duration which prolongs with
increasing $\beta$. Afterwards the most dominant natural orbital loses weight
in favour for the second dominant one, which happens much faster for the black
soliton compared to $\beta=0.5$ on the
time-scales under consideration.

\begin{figure}[t]
 \centering
\includegraphics[width=0.5\textwidth]{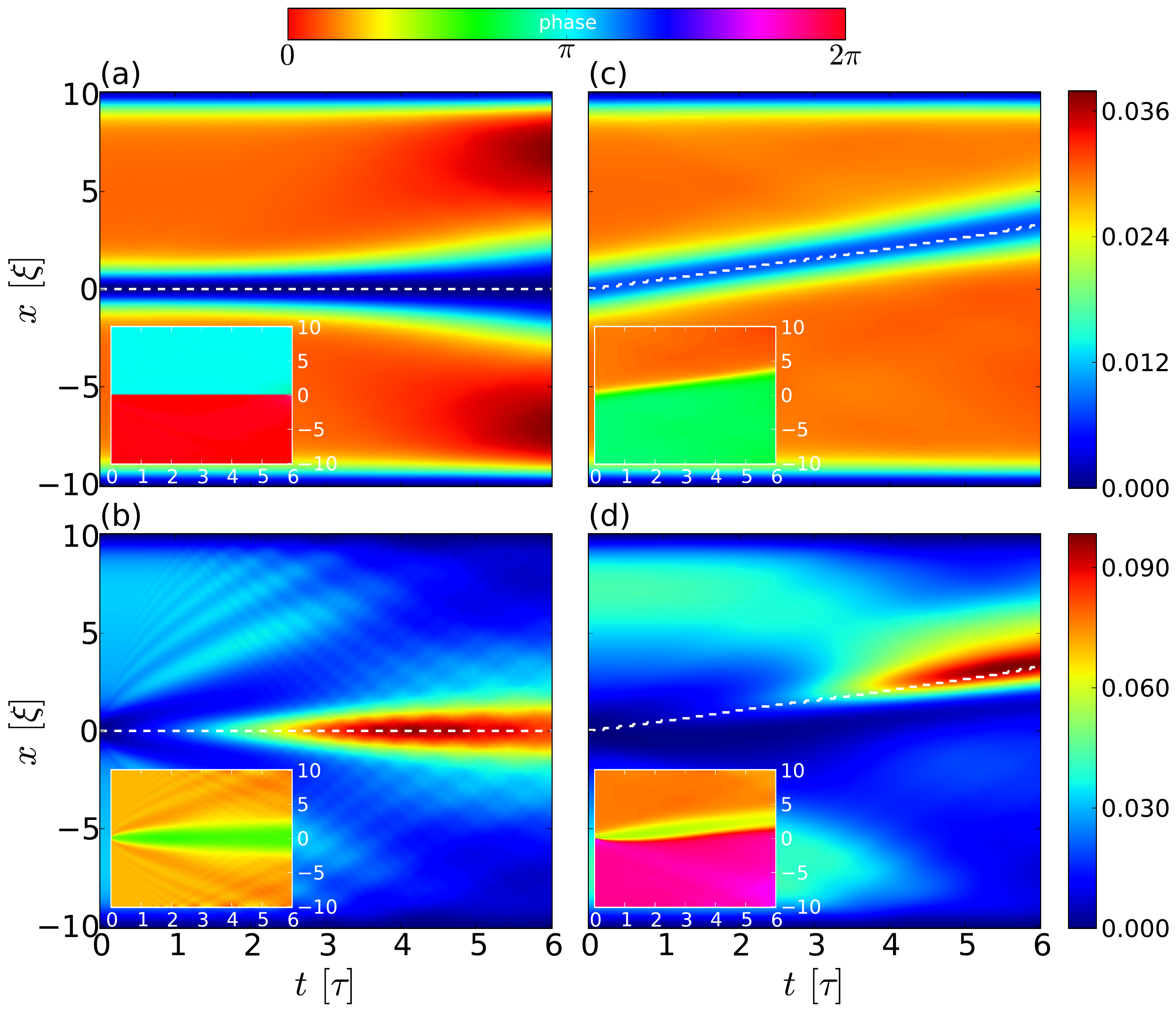}
\caption{Comparison of the density evolution of the most and second dominant NO
for a
black soliton ((a) and (b)) and a grey soliton with $\beta=0.5$ 
((c) and (d)). The instantaneous soliton positions $x^s_t$ defined as the
minimum of 
$\rho_1(x;t)$ are depicted as white dashed lines.
Insets: Phase profile of respective natural orbital.
All other parameters as in Fig.\ \ref{fig_density}.}
\label{fig_norbs}
\end{figure}
Finally, we unravel the density- and phase-profiles of the two most dominant
natural orbitals in Fig.\ \ref{fig_norbs}. Clearly, the dominant natural orbital
features solitonic
characteristics such as the localized density minimum accompanied with an
appropriate phase jump. In contrast to this, the second dominant natural orbital undergoes
two stages of evolution. Firstly, the density is rearranged on a time-scale, 
on which also the depletion and the natural populations are
stationary, i.e. until $t\sim1.5\tau,...,3.5\tau$ for $\beta=0.0,...,0.7$.
After this phase of the dynamics, the second dominant natural orbital has
accumulated most of its density in the vicinity of the instantaneous density
minimum of the dominant natural orbital. This accumulated density remains localized 
in the vicinity of the density dip of the dominant natural orbital even in the case of
a moving grey soliton. In view of its natural population and density distribution, the second
dominant natural orbital is mainly responsible for the filling of the one-body density
depression -
a result fully consistent with 
lattice simulations for black solitons in a box potential
\cite{quantum_mb_dynamics_of_dark_solitons_in_optical_lattices_Carr_PRA09} as
well as a Bogoliubov treatment of black solitons in a harmonic trap, where the
anomalous mode dominates the filling of the density minimum, e.g.
\cite{quantum_depl_dark_sol_anomalous_mode2002,law_quantum_2002}.

In Fig.\ \ref{fig_norbs}, we have moreover marked the position of the soliton
allowing for identifying a crucial difference between black and grey solitons:
For $\beta=0.0$, the many-body state has a well defined parity for all times,
which translates into well defined parities of the natural orbitals and therefore leads to
a perfectly symmetrically accumulated density of the second dominant natural orbital with
respect to the soliton position. In contrast to this, the many-body state for
$\beta>0$ can be shown to feature only a combined parity and
time-reversal symmetry and,
more importantly, the accumulated density of the second dominant natural orbital is
not {\it locally} symmetric with respect to the instantaneous soliton position.
In Sect.\ \ref{sec_g2}, we will see that this asymmetry results in distinct
two-body correlations. We remark that the degree of local
asymmetry depends of the definition of the soliton position $x^s_t$: If we had
defined $x^s_t$ as the minimum of the dominant natural orbital density
$|\varphi_1(x;t)|^2$, the local asymmetry would be still present but slightly
reduced. All statements in Sect.\ \ref{sec_g2} and \ref{sec_no_impr}, 
which relate to the soliton 
position, hold qualitatively for both definitions of $x^s_t$ and we
stick to $x^s_t$ being the position of the 
minimum of $\rho_1(x;t)$ as the latter is
directly observable.

Before turning to the local two-body correlations, we comment on phonon
excitation mechanisms: In the case of
the black soliton, we observe that the even parity natural orbitals, i.e. the
second and fourth one (not shown), are essentially carrying all phonon
excitations visible in $\rho_1(x;t)$. These natural orbitals have been of odd parity before
the phase-imprinting and possess a finite slope in the vicinity of $x=0$. Thus
the phase-engineering creates a cusp at $x=0$, whose energy density is
subsequently transported via phonons into the bulk as one can infer from
the oscillatory density- and phase-modulations in Fig.\ \ref{fig_norbs} (b). 
Yet also the odd parity natural orbitals
contribute to the phonons in $\rho_1(x;t)$ with, however, so minute weight that
the density modulations are hardly observable in Fig.\ \ref{fig_norbs} (a). Here
the phonon generation underlies a different mechanism: Having been of even
parity before the phase-imprinting, odd parity natural orbitals initially feature a tiny
but finite density at the phase-jump position $x=0$, which is transported into
the bulk afterwards.  It is conceivable that
besides these two
phonon generation mechanisms also the shape of $|\varphi_i(x;t)|^2$ in a finite
vicinity of $x=0$ plays a role by 
storing excess energy
when the density-engineering barrier is removed, which is then turned into
phonon excitations. However, this argument should generically hold for all
natural orbitals irrespectively of their parity and therefore we may regard this
mechanism to be of minor importance if it is important at all.
For $\beta=0.5$,
phonons are hardly visible in the natural orbitals and therefore essentially absent in
$\rho_1(x;t)$ as already discussed.
\subsection{Two-body correlation analysis}\label{sec_g2}
As solitons are entities localized in space, we consider the local two-body
correlation measure:
\begin{equation}\label{def_g2}
 g_2(x_1,x_2;t)=\frac{\rho_2(x_1,x_2;t)}{\rho_1(x_1;t)\rho_1(x_2;t)},
\end{equation}
which coincides with the diagonal of the second order coherence function
defined by Glauber \cite{glauber_quantum_1963}. Here, we have introduced the
two-body density $\rho_2(x_1,x_2;t)=\langle x_1x_2|\hat\rho_2(t) |x_1x_2\rangle$
with the reduced two-body density operator
$\hat\rho_2(t)={\rm tr}_2|\Psi_t\rangle\!\langle \Psi_t|$ obtained by a partial
trace
over all but
two bosons. Due to our normalization of $\rho_2(x_1,x_2;t)$ to unity, a
perfectly
condensed many-body state
would lead to
$g_2(x_1,x_2;t)=1$ everywhere. Finding $g_2(x_1,x_2;t)$ to be larger (smaller)
than unity means that two bosons are found more (less) likely at the spots
$(x_1,x_2)$ compared to statistical independence. Despite of the
$g_2$-correlation function being one of the simplest observables
sensitive to beyond mean-field properties, such correlations have not been
studied in the context of dark solitons - except for the perturbative
treatment with a different focus in
\cite{quantum_fluctuations_in_image_of_BEC_NegrettiPRA08}.
Besides its conceptual simplicity, the experimental accessibility via {\it in
situ}
density-density fluctuation measurements
\cite{hung_extracting_2011,endres_observation_2011,
cheneau_light-cone-like_2012,endres_single-site-_2013,
jacqmin_sub-poissonian_2011,armijo_mapping_2011,armijo_direct_2012} makes the
$g_2$-correlation
function attractive
such that experimental interest in the correlation properties of
dark solitons has already been triggered \cite{armijo_direct_2012}. 
Due to the involved averaging process, probing beyond 
mean-field physics on the level of density fluctuations
can turn out to be more robust than inspecting signatures becoming
manifest only in single-shot $n$-body ($n\gg 2$) absorption image measurements
as considered in e.g.\ \cite{delande_many-body_2014} - given a sufficient 
stability of the apparatus and imaging
system, of course.	

\begin{figure*}[t]
 \centering
\includegraphics[width=0.9\textwidth]{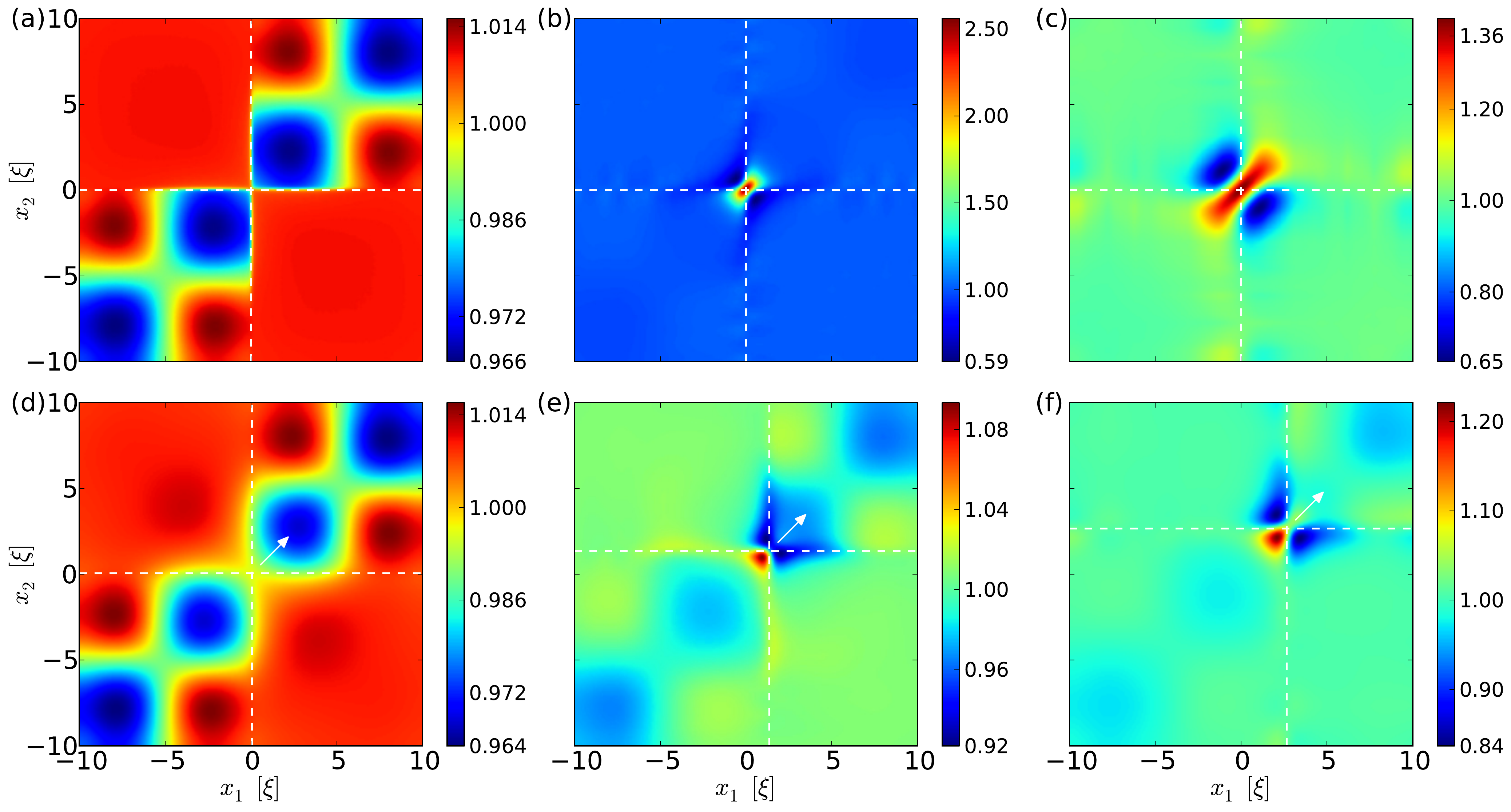}
\caption{Two-body correlation function $g_2(x_1,x_2;t)$ for a black soliton
(first row) and
a grey soliton with $\beta=0.5$ (second) at times $t=0.0$ (first column),
$t=2.5\tau$ (second) and $t=5\tau$ (third). Dashed lines refer to 
the instantaneous soliton position $x^s_t$ and the arrows indicate the
direction of the soliton movement.
All other parameters as in Fig.\ \ref{fig_density}.}
\label{fig_g2}
\end{figure*}

In Fig.\ \ref{fig_g2}, we compare the time evolution of the $g_2$-correlation
measure	
for a black and a grey soliton initial state: Initially (Fig.\ \ref{fig_g2}
(a), (d)), $g_2(x_1,x_2;t=0)$
only marginally deviates from unity as expected for an initial state within
the Thomas-Fermi mean-field regime. During the time evolution, however, the
black and grey soliton establish quite distinct correlation patterns:
Firstly focussing on the black soliton, the $g_2$-correlation function inherits
the two-body reflection symmetry $g_2(-x_1,-x_2;t)=g_2(x_1,x_2;t)$ from the
many-body wave function of well defined parity. Most of the detection events
$(x_1,x_2)$ remain uncorrelated during the evolution, but pronounced
correlations emerge in the vicinity of the soliton notch (Fig.
\ref{fig_g2}
(b)): 
On the one hand side, we observe that a pair of bosons strongly bunches
at the soliton position ($x_1=x_2=0$) or in the same flank of the soliton
($x_1x_2>0$).  On the
other hand, two bosons statistically avoid to be detected in different flanks
of the soliton ($x_1x_2<0$). Thereby, the region around the soliton position where
$g_2(x_1,x_2;t)$ features bunching is strongly squeezed on the off-diagonal
$x_2=-x_1$ axis.
At later times (Fig.\ \ref{fig_g2} (c)), the $g_2$-function
preserves its correlation pattern but its maximal value reduces by more than a
factor $1.8$ compared to (Fig.\ \ref{fig_g2} (b)). At the same time, the
extent of the bunching (anti-bunching) region on the diagonal $x_1=x_2$
(off-diagonal $x_1=-x_2$) increases from about $1.7\xi$ ($1.0\xi$) to $3.0\xi$
($2.2\xi$), which goes hand in hand with the widening of the minimum in the
reduced one-body density (cf.\ Fig.\ \ref{fig_density} (a) and \ref{fig_norbs} 
(a)).

In the case of a grey soliton, the broken parity symmetry is imprinted in
the evolution of the $g_2$-correlation function leading to
quite a characteristic correlation pattern after a short time (Fig.
\ref{fig_g2} (e)): Most strikingly, the region where atom pairs occur bunched 
is localized in the soliton flank opposite to its direction of movement, i.e.
$x_1,x_2<x^s_t$, whereas atoms in the
soliton flank pointing into its direction of motion are either slightly
anti-bunched or uncorrelated. 
In fact, this displacement of the bunching region can be - at least partially -
traced back to the local asymmetry of the second dominant natural orbital density w.r.t.
the soliton position $x^s_t$: Focussing
on the diagonal $x_1=x_2=x^s_t\pm\epsilon$, $\epsilon > 0$ 
and on the vicinity of $x^s_t$ such that 
$\rho_1(x^s_t\pm\epsilon;t)\approx \lambda_2(t)|\varphi_2(x^s_t\pm\epsilon;t)|^2$
(cf.\ Fig.\ \ref{fig_norbs} (c)),
one immediately sees that,
for fixed two-body densities 
$\rho_2(x^s_t\pm\epsilon,x^s_t\pm\epsilon;t)$, the local asymmetry 
$|\varphi_2(x^s_t-\epsilon;t)|^2<|\varphi_2(x^s_t+\epsilon;t)|^2$ 
implies
$g_2(x^s_t+\epsilon,x^s_t+\epsilon;t)<g_2(x^s_t-\epsilon,x^s_t-\epsilon;t)$
according to the definition (\ref{def_g2}).

Moreover, regions of 
anti-bunching emerge in the sectors $(x_1,x_2)\in\mathbb{R}^2$ with 
$x_{1/2}<x^s_t<x_{2/1}$, which
are elongated towards the direction of motion of the soliton.
In the overall correlation pattern, the maximal deviation of $g_2(x_1,x_2;t)$
from unity
is with $\pm 8\%$ 
rather weak but increases 
at later times (Fig.
\ref{fig_g2} (f)). Furthermore, the anti-bunching regions become confined to
the sectors $x_{1/2}<x^s_t<x_{2/1}$ then while an atom pair in the sector 
pointing into the soliton's direction of
movement, i.e. $x_1,x_2>x^s_t$, turns out to be 
essentially
uncorrelated. At even later times, the bunching region
widens and shifts in the direction of movement such that it becomes
approximately symmetric on the diagonal $x_1=x_2$ with respect to the soliton
position $x^s_t$ (plot not shown). We note that the localization of the (anti-) 
bunching regions with respect to the soliton position does not depend on
the choice for the definition of $x^s_t$ discussed in Sect.\ 
\ref{ssec_dep_norb}.

In total, the maximal
positive and negative deviations of $g_2(x_1,x_2;t)$ from unity are much
smaller compared to the black soliton, which can be easily understood in terms
of selection rules: 
While for $\beta=0$, the parity symmetry of the total wave function allows 
only to scatter atoms pairwise out of the most
dominant natural orbital of odd
parity into the even parity second most dominant NO
(cf.\ also
\cite{
quantum_depletion_of_excited_condensate_Sacha_Dziarmaga_Karkuszewski_PRA2002,
quantum_entangled_dark_solitons_in_optical_lattices_Carr_PRL09,
quantum_mb_dynamics_of_dark_solitons_in_optical_lattices_Carr_PRA09}),
this process can also happen atom-wise in the case of a grey soliton. In view
of the particular density distribution of the most and second most dominant
natural orbital, two-body correlations have thereby to be quite pronounced for the black
soliton, whereas relatively weaker correlations are possible for grey solitons.

\begin{figure}[t]
 \centering
\includegraphics[width=0.45\textwidth]{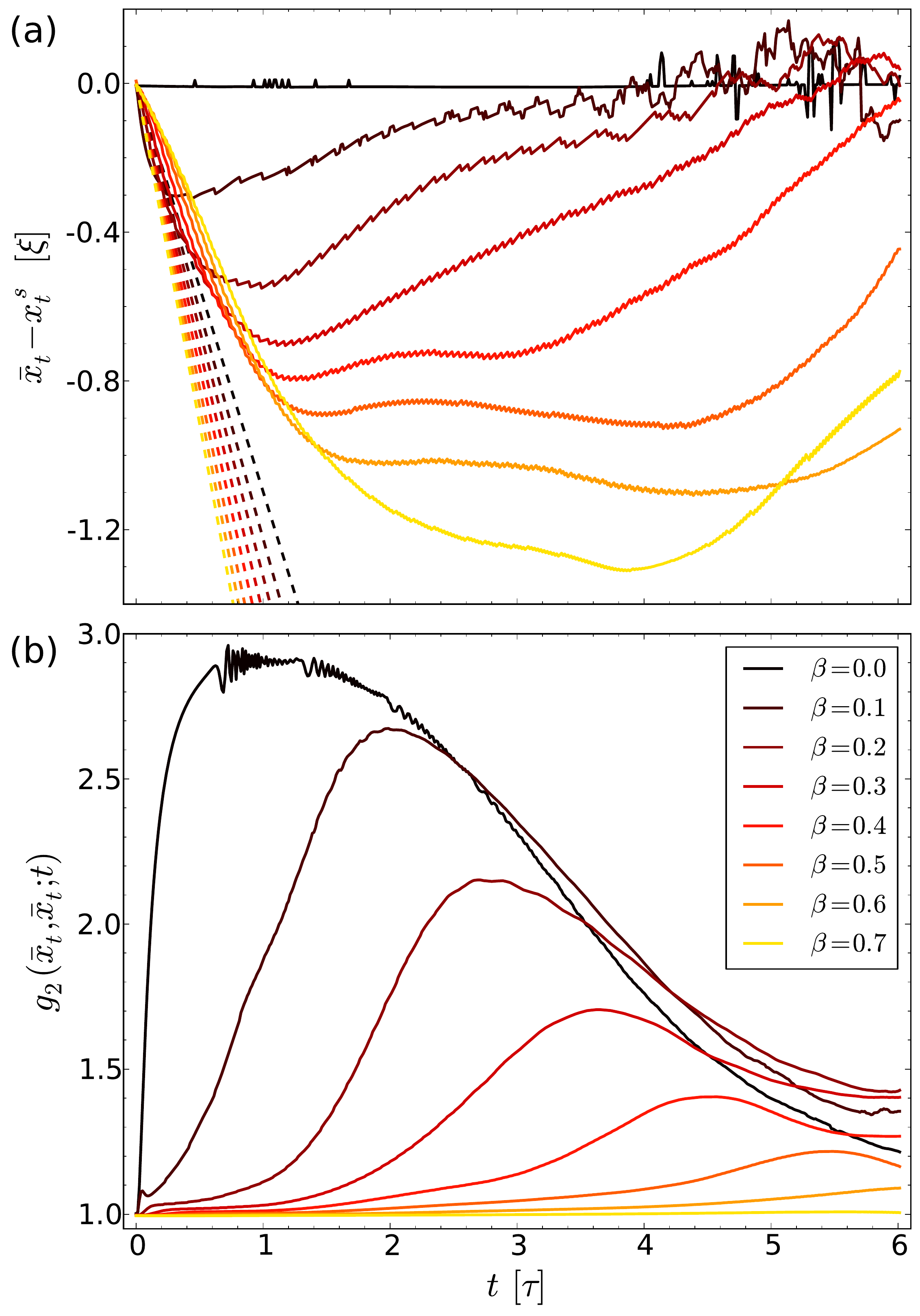}
\caption{Analysis of the bunching centre $\bar{x}_t$ for various $\beta$:  (a)
Displacement of 
$\bar{x}_t$ with respect to instantaneous soliton position $x^s_t$. The 
dashed lines indicate the trajectories of a fictitious sonic excitation 
emitted at $t=0$ from the soliton position in negative $x$-direction for the
various $\beta$, i.e. 
$x(t)-x^s_t=-\sqrt{g\rho_{\rm bulk}}\,t-x^s_t$, where
the bulk density $\rho_{\rm bulk}$
is evaluated directly after the density-engineering at $t=0$.
(b) The
value of the two-body correlation function $g_2(\bar{x}_t,\bar{x}_t;t)$ at 
the bunching centre. The bunching distribution (\ref{def_bunch_distr}) is
calculated for $R=4\xi$. All other parameters as in Fig.\ \ref{fig_density}.}.
\label{fig_bunch_center}	
\end{figure}

In order to quantify the asymmetry of the correlation pattern in dependence 
on $\beta$, we define the position of the bunching centre as follows: Firstly,
we introduce a bunching distribution within a disk of radius $R$ with centre 
$(x^s_t,x^s_t)$  as being proportional to the $g_2$-function 
minus one wherever it shows bunching:
\begin{eqnarray}\label{def_bunch_distr}
 p(x_1,x_2;t) \propto
&\phantom{x}&\left[\,g_2(x_1,x_2;t)-1\,\right]\;\times\\\nonumber
&\phantom{x}&\times\; \Theta(R-|{\bf x}-{\bf x}^s_t|)\;
\Theta(g_2(x_1,x_2;t)-1).
\end{eqnarray}
Here, $\Theta(x)$ denotes the Heaviside step function and we use the 
abbreviations ${\bf x}=(x_1,x_2)$ as well as
${\bf x}^s_t=(x^s_t,x^s_t)$. In the following, we choose $R=4\xi$, which is
sufficient for capturing the important correlation pattern. 
This probability distribution 
is used for defining the bunching centre
${\bf \bar x}_t$ as the expectation value of ${\bf x}$.
Due to the particle exchange symmetry of the $g_2$-correlation function, both
components of ${\bf \bar x}_t$ coincide and equal:
\begin{equation}\label{def_bunch_center}
 \bar{x}_t=\int {\rm d}x_1{\rm d}x_2\,x_1\, p(x_1,x_2;t).
\end{equation}
For comparing the $\bar{x}_t$ dynamics for 
various $\beta$, we have performed a Galilean boost into the co-moving system of
the soliton in Fig.\ \ref{fig_bunch_center} (a) depicting $\bar{x}_t-x^s_t$,
which is proportional to bunching centre displacement from the soliton position
in the $(x_1,x_2)$-plane, i.e. $|{\bf \bar x}_t-{\bf
x}^s_t|=\sqrt{2}|\bar{x}_t-x^s_t|$.
Within $0.5,...,2.5$ natural time units $\tau$ for
$\beta=0.1,...,0.7$, 
the bunching centre
$\bar{x}_t$ firstly moves away from the soliton opposite to its direction of
motion with an approximately constant velocity being the faster the smaller
$\beta$ (for $\beta>0$). This motion takes place with subsonic
velocity\footnote{We remark that
for $\beta=0.1$ and $\beta=0.2$ the motion appears to be supersonic. Since the
bunching centre, however, is only slightly displaced from the soliton position
for these slow solitons, a comparison with the bulk sound velocity turns out to
be difficult: The bunching centre stays in a region of very low and spatially
rapidly changing density such that the local speed
of sound (which actually is not a meaningful concept here in view of the spatial
extent of the bunching region) turns out to be much smaller than its bulk value.
See also the discussion about event horizons and dark solitons in
\cite{walczak_exact_2011}.} as one can infer from
a comparison with the trajectory $x(t)$ of a fictitious sonic excitation emitted
opposite
to the soliton direction of movement at $t=0$ 
from the soliton position $x(0)=0$.
Importantly, this
steady motion of the bunching centre lasts the longer the greyer the soliton is 
such that larger
$\beta$ result in larger maximal displacements from the soliton position. After
this phase of the dynamics, the displacement either features a maximum or stays
for some time approximately at its maximal value before it decreases. This
decrease reflects that the bunching centre becomes approximately symmetric with
respect to the soliton position at even later times. Afterwards, we cannot make
predictions for the further evolution since more optimized SPFs would be needed
for ensuring convergence.

Although the term is definitely not proper in a strict sense, these findings 
suggest that the
observed highly localized correlations feature a certain ``inertia'': The
phase-imprinting appears to give the density dip an instantaneous kick setting
it thereby into motion. At the same time, bunching correlations emerge and
drift the farer into the
back of the soliton the
stronger this kick is, i.e. the larger $\beta$. In order to test this picture,
we have exerted a force on a grey soliton giving it dynamically a kick: By
imposing
an additional weak potential $V(x)=V_0\,[1-\tanh(x/l)]/2$ with $V_0>0$, 
$l=\mathcal{O}(\xi)$, we realize
a bulk density profile $\propto [{\rm const.} - V(x)]$ featuring two
distinct sound velocities $s_l < s_r$ in the left /right half space. We then
initialize a grey soliton in the left half space ($x_0<0$ in
(\ref{grey_sol_analy}))
moving to the right 
by means of the optimized density- and phase-engineering scheme introduced in
Sect.\ \ref{ssec_initial_state}. Although $l=\mathcal{O}(\xi)$
results in a non-adiabatic change of the local density for the soliton (cf.
e.g.\ 
\cite{motion_dark_solitons_in_trapped_BEC_Busch_PRL2000}), we have carefully
checked
that passing the step in the bulk density only leads to an 
acceleration of the soliton within the mean-field picture. The corresponding 
many-body simulation reveals that the bunching centre
becomes drastically separated from the soliton position when passing $x=0$.
Essentially, the bunching centre remains stuck in the vicinity of $x=0$ while 
decreasing in amplitude with time, giving thus further evidence for 
the aforementioned ``inertia''
of these localized correlations under kicks 
(plots not shown). We suspect that this ``inertia'' effect
might be a consequence of the time-scales for the emergence
and drift of these correlations being decoupled from the time-scales
associated with the movement and acceleration of the density dip.

In Fig.\ \ref{fig_bunch_center} (b), we present the time evolution of the
$g_2$-correlation function evaluated at the bunching centre, i.e.
$g_2(\bar{x}_t,\bar{x}_t;t)$. For the black soliton, strong bunching
correlations emerge almost instantaneously, which reflects the particular role
of the parity induced selection rule, enforcing the 
dominant dynamical quantum depletion channel to take place pairwise,
as well as the particular shape of the
dominant and second dominant natural orbital for local two-body correlations. 
At about $t=1.0\tau$, the bunching correlations establish a maximum of
approximately $2.9$ and afterwards these correlations decay again. We conjecture
that this decay might be a precursor of a relaxation to a stationary state at 
later times. Qualitatively, the
$g_2(\bar{x}_t,\bar{x}_t;t)$ follows essentially the same behaviour in the case
of a grey soliton. However, the faster the soliton the longer does it take until
noticeable correlations have built up. The build-up of correlations can last
several natural time units $\tau$, i.e. is significantly longer compared to
the black soliton. Moreover, the time when
$g_2(\bar{x}_t,\bar{x}_t;t)$ becomes maximal turns out to be longer than the
time needed for the bunching centre becoming maximally displaced from the
soliton position.
As expected from the previous observations, the maximum of
$g_2(\bar{x}_t,\bar{x}_t;t)$ is smaller for larger $\beta$.
\subsection{Role of phase-engineering}\label{sec_no_impr}
\begin{figure*}[t]
 \centering
\includegraphics[width=0.75\textwidth]{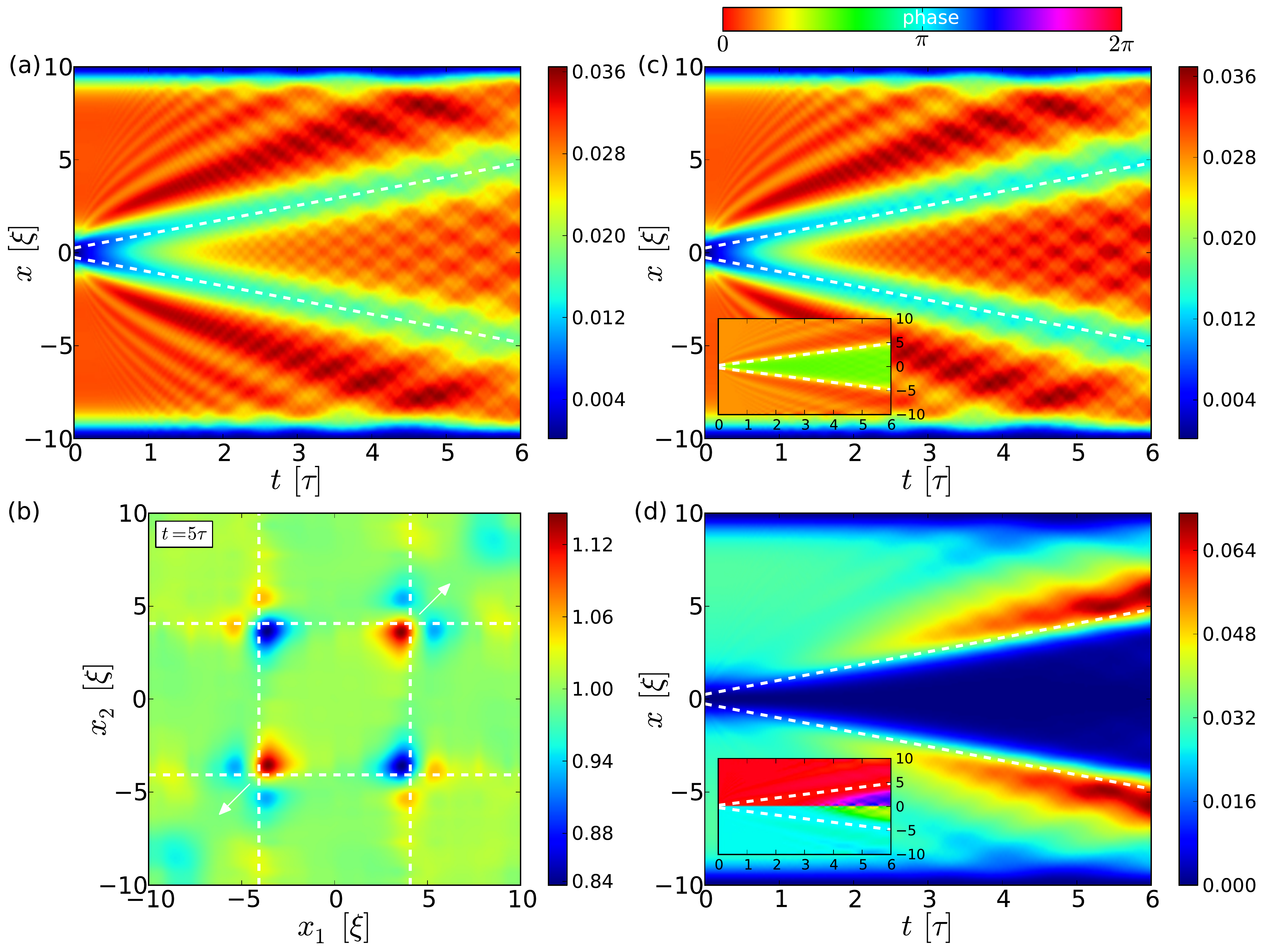}
\caption{Evolution of a density-engineered initial state 
(no phase-imprinting is applied):
Reduced one-body density $\rho_1(x;t)$ (a), two-body correlation function
$g_2(x_1,x_2)$ 
at time $t=5\tau$ (b) as well as density of the dominant (c) and second
dominant natural orbital (d) with corresponding phase profiles as insets. The
corresponding
natural
populations are depicted in Fig.\ \ref{fig_depl_npop} (b).
Dashed lines refer to 
the instantaneous soliton position $x^s_t$ obtained by linear regression
of the position of the local $\rho_1(x;t)$ minima and the arrows indicate the
direction of the soliton movement.
All other parameters as in Fig.\ \ref{fig_density}.}
\label{fig_no_impr}
\end{figure*}
Having discussed the dynamics of a phase- and density-engineered solitonic
initial state in detail, we finally investigate the role of the
phase-imprinting procedure. For this purpose, we only apply the 
density-engineering scheme for a black soliton\footnote{Preparing the initial
state with the
optimized density-engineering potential (\ref{dens_eng_pot}) for $\beta>0$ 
instead does not
change the results qualitatively. Yet as seen for the phase- and
density-engineered initial state, the strength of beyond mean-field effects such
as
correlations are weaker in this case. We note that this density-engineering
scheme
results in a state of well-defined many-body parity also for $\beta>0$.} as 
described in Sect.\ \ref{ssec_initial_state}.	
As a result, we obtain an initial state with the very same natural population 
distribution as
in the case of phase- and density-engineering. In contrast to the latter
situation, 
the majority of atoms resides in a gerade rather than an ungerade orbital now, 
which constitutes an energetically more favourable situation\footnote{In fact,
the energy of the many-body system is enhanced by the phase-engineering 
not only due to altered parity of the orbitals but also because of the fact that
a phase step is imprinted in a region of finite density.
Comparing the excess energy of the density-engineered with the density- and
phase-engineered initial state, i.e. $E_{\rm d}-E_0$ and $E_{\rm dp}-E_0$
with $E_0$ denoting the ground state energy of the considered $N=100$ bosons
in the box
potential,
we find $(E_{\rm dp}-E_0)/(E_{\rm d}-E_0) \approx 1.09$.}. The
subsequent
many-body dynamics is summarized in Fig.\ \ref{fig_no_impr}.

In contrast to the phase- and density-engineered initial state, the density
minimum now splits into two counter-propagating minima of constant velocities,
which move through a background with phonon modes being excited much more 
intensively (Fig.\ 
\ref{fig_no_impr} (a)). A mean-field simulation (plot not shown) essentially  
reveals the same density profile $\rho_1(x;t)$, which can be explained by the
evolution of the natural populations in Fig.\ \ref{fig_depl_npop} (b): Quite in
contrast to
the density- and phase-engineered initial state, the natural populations essentially stay 
constant and, thus, the rather small initial depletion of about $8.5\%$ is 
retained. The density of the dominant natural orbital approximately resembles the full
density profile $\rho_1(x;t)$ as expected while its phase profile 
features two phase jumps localized at the positions of these two minima (Fig.
\ref{fig_no_impr} (c)). Therefore, we may conclude that the density-engineering
scheme creates two counter-propagating grey solitons with $|\beta|\approx 0.7$
in the
single-particle state occupied by approximately $91.5\%$ of the atoms - as
predicted for density-engineering 
in \cite{gredeskul_generation_1989-1,gredeskul_dark-pulse_1990-1}
within the mean-field theory and also experimentally observed in e.g.
\cite{dutton_observation_2001}. 

The remaining atoms essentially reside in the
second dominant natural orbital being of odd parity, whose density appears as if being
dragged to the box boundaries by the two grey solitons in the dominant natural orbital.
Thereby, the region between the two solitons becomes depleted in density while
density is accumulated in vicinity of the two density minima of the dominant
natural orbital. Again, we can witness a local asymmetry of this accumulated density with
respect to the position of the two grey solitons in the dominant natural orbital, which
is a persistent feature for both definitions of the soliton position 
discussed in Sect.\ \ref{ssec_dep_norb}. Moreover, we note that the
phase of the second dominant natural orbital is approximately constant in domains
$x\in[-L/2,0)$ and $x\in(0,L/2]$ so that this orbital hardly contributes 
to the probability current density $j(x;t)={\rm tr}(\hat
\jmath(x) \hat\rho_1(t))$ with the current density operator 
$\hat \jmath(x) = \frac{1}{2}[\delta(x-\hat x)\,\hat p + \hat p\,\delta(x-\hat
x)]$
and $\hat x$, $\hat p$ denoting the position and momentum operator,
respectively. 
Thus, essentially only the phonon excitations being almost
exclusively accommodated in the gerade dominant natural orbital as well as the mass
counter-flow to the movement of the grey solitons in this orbital contribute to
the current density $j(x;t)$.

Summarizing, all observations so far are consistent with our results on a
single grey soliton, in particular the long lifetime of the contrast against
decay via quantum fluctuations (cf.\ Fig.\ \ref{fig_contrast} (b), $\beta=0.7$).
We finally address the question in Fig.\ \ref{fig_no_impr} (b) whether also the
$g_2$-correlation pattern
can be understood as the sum of the correlation patterns of two
counter-propagating grey solitons: Indeed, we find for $x_1$, $x_2$ in
the vicinity of one and the same grey soliton that 
$g_2(x_1,x_2;t)$ resembles the peculiar locally asymmetric correlation pattern
of bunching in the back of the soliton, anti-bunching of pairs of atoms being
located in different flanks of the soliton and being uncorrelated ahead. Yet
moreover, an additional, with respect to the soliton position non-local
correlation structure has emerged. Denoting the position of e.g.\ the right
moving grey soliton as $x^s_t$, we may state that bunching (anti-bunching)
regions for $(x_1, x_2)$ in the vicinity of $(x^s_t,x^s_t)$ are turned into
anti-bunching (bunching)
regions under a parity transformation acting only on one of the two coordinates
$(x_1,x_2)$: Finding one atom in each of the backs
of the two solitons is statistically avoided while detecting one atom in each
of the forward flanks of the two solitons is a rather uncorrelated event.
Moreover, the probability of measuring one atom in the back flank of one
soliton and another atom in the forward flank of the other soliton is slightly
enhanced due to correlations. As a matter of fact, these non-local
correlations between the solitons are a generic feature of parity symmetric
many-body wave functions with essentially only two contributing
natural orbitals and $g_2(0,0;t)\approx 1$ \cite{kronke_two-body_2015}.
\section{Conclusions and outlook}\label{sec_concl}
We have provided a systematic study of beyond mean-field
signatures in black and, in particular, grey solitons. For this purpose, a
robust, semi-analytically optimized density-engineering scheme has been
developed, which in combination with phase-imprinting allows to cleanly generate
grey solitons. In situations where the healing length can be increased above
the optical diffraction limit, this preparation scheme is also of potential
experimental relevance. We have demonstrated that the quantum
fluctuations limited lifetime of dark solitons increases with their velocity.
This is, in particular, intriguing as the variety of allowed incoherent
scattering channels is larger for grey solitons compared to black ones of
well defined party. The enhanced lifetime of grey solitons also manifests
itself in a slower dynamical quantum depletion. The dark soliton decay takes
place
in a two step process: Firstly, the density of the second dominant
natural orbital accumulates in the vicinity of the soliton while the depletion
stays constant. Secondly, the population of this orbital increases
significantly. Strikingly, the accumulated density of the second dominant
natural orbital features a local asymmetry w.r.t. to the soliton position for
grey solitons. 

Dark solitons have the unique feature that quantum 
fluctuations induce spatially highly localized two-body correlations 
in the vicinity of the density minimum. While the zones of (anti-)bunching are
distributed symmetrically around this minimum for a black soliton of well 
defined parity, the locally asymmetrically accumulated density of the second 
dominant 
natural orbital imprints itself in an asymmetric correlation pattern for grey 
solitons. In particular, we have shown that localized bunching correlations 
move to the backward flank of a grey soliton with subsonic velocity resulting
in a bunching centre being the farer displaced from the soliton the faster the
soliton moves through the bulk. Moreover, we have observed that these localized
correlations have kind of a particle character in the sense that they feature
a certain inertia under accelerations of the soliton. To unravel the
underlying mechanism and sharpen the terminology for this phenomenology 
remains an interesting prospect for future works. 

Finally, we have illuminated the role of the phase-imprinting: As within 
the mean-field approximation, density engineering alone results in pairs of
counter-propagating grey solitons, which individually feature both the 
enhanced lifetime and peculiar localized correlation pattern of 
a single grey soliton. In addition, non-local two-body correlations
between 
the two solitons emerge, which can be traced back to the parity symmetry,
absent correlations at the parity-symmetry centre, $g_2(0,0;t)\approx1$, and
the fact that essentially only two natural orbitals contribute with
significant weights \cite{kronke_two-body_2015}.
As a next step, it would be interesting to also study dark-bright
solitons beyond the mean-field approximation in order to reveal possibly 
emerging inter-species correlations. Moreover, it is necessary to check
the robustness of all these beyond mean-field signatures in the 
presence of decoherence and particle loss, which has been shown to
significantly influence the properties of quantum bright matter wave
solitons \cite{weiss_short-time_2014}.

\begin{acknowledgments}
We would like to thank Krzysztof Sacha, Dimitri Frantzeskakis,
 Vassos Achilleos, Panayotis Kevrekidis, Antonio Negretti as well as
 Juliette Simonet for
inspiring discussions
 and Dominique Delande for providing natural population
data of the TEBD simulations from \cite{delande_many-body_2014}. 
S.K. gratefully acknowledges a scholarship by the Studienstiftung des deutschen
Volkes. P.S. gratefully acknowledges financial support by the
Deutsche Forschungsgemeinschaft in the framework of the project
\mbox{Schm 885/26-1}.
\end{acknowledgments}
\appendix
\section{Numerical parameters and convergence}\label{app_convergence}
\setcounter{section}{1}
The hard wall boundary conditions are implemented by a sine discrete variable
representation (DVR) with $n=200$ grid points resulting in a grid
spacing
$\Delta x\approx0.1\xi$ (cf.\ appendix of \cite{MCTDH_BJMW2000}). 
Compared to the Bose-Hubbard approximation of a
continuous system (cf.\ e.g.\
\cite{BH_cont_space_Fleischhauer_PRA2007}), 
our sine DVR also considers next-to-nearest neighbour and
higher order hopping processes. Due to the single-particle spectrum
of the density-engineering Hamiltonian featuring pairs of quasi-degenerate
states, it is only meaningful to consider an even number of SPFs when going
beyond the mean-field approximation. We can efficiently afford for at most
$M=4$ optimized SPFs when dealing with $N=100$ bosons resulting in $\sim177.000$
number state configurations. By carefully comparing with $M=2$
simulations as well as with the results in \cite{delande_many-body_2014},
we can ensure convergence at least up to times $t\sim 6.25\tau$, 
which
is not long enough for relaxing to a uniform density profile but more than
sufficient for the phenomena we are interested in. 
We emphasise that (ML-)MCTDHB gives a variationally optimized total
wave function for any $M$.
\bibliography{soliton1_no_title}
\end{document}